\documentclass{cimento}
\usepackage{graphicx}

\title{Selected Items in Jet Algorithms}
\author{Giuseppe Bozzi\from{ins:1}}
\instlist{\inst{ins:1}Institut f\"ur Theoretische Physik, Universit\"at Karlsruhe, P.O.Box 6980, 76128 Karlsruhe, Germany}
\PACSes{\PACSit{13.87.A}{Jets in large-$Q^2$ scattering}}

\begin{document}

\maketitle

\begin{abstract}
I provide a very brief overview of recent developments in jet
algorithms, mostly focusing on the issue of infrared-safety.
\end{abstract}

\section{Introduction}

The main properties that should be satisfied by any jet definition were
already pointed out almost twenty years ago
\cite{Ellis:1989vm,Huth:1990mi}: the jet algorithm should be 1) simple to
implement both in experimental analysis and theoretical calculations, 2)
well defined and yielding a finite cross section at any order of
perturbation theory, 3) relatively insensitive to hadronization.
Many different jet definitions have been proposed in recent years
but it turns out that some of them do not strictly meet the above
features. This could in principle lead to serious problems
especially when infrared(IR)-safety (i.e., the second item in the above list)
is not correctly implemented, since in this case the matching with
fixed-order theoretical results would be spoiled and the whole
jet-finding procedure would heavily depend on non-perturbative effects
(hadronization, underlying event, pile-up). 
We can group jet algorithms in two broad classes: Iterative Cone (IC) and
Sequential Recombination (SR). IC algorithms cluster particles according
to their relative distance in {\it coordinate-space} and have been
extensively used at past lepton and hadron colliders. SR algorithms
cluster particles according to their relative distance in {\it
  momentum-space} and are somehow preferred by theorists since they
rigorously take into account IR-safety. In what follows I will try to
give an overview of recent developments obtained in both the IC and SR
classes. For a complete and extensive treatment of these and other
aspects of jet algorithms I refer the reader to the recent literature on
the subject \cite{Ellis:2007ib,Buttar:2008jx}.   

\section{Iterative-Cone algorithms}

The core structure of any IC algorithm \cite{Arnison:1983rn} can be
described as follows: choose a {\it seed} particle {\it i}, sum the momenta of
all particles {\it j} within a cone of radius {\it R} (in {\it y} and
 $\phi$) around {\it i}, take the direction of this sum as a new seed,
 and repeat until the cone is stable (i.e., the direction of the 
{\it n}-th seed coincides with the direction of the {\it (n-1)}-th).
This procedure, however, may eventually lead to find multiple stable
cones sharing the same particles, i.e. the cones may overlap. The
problem can be solved either by preventing the overlap (Progressive
Removal (IC-PR) algorithms) or through a splitting procedure (Split-Merge
(IC-SM) algorithms).
The IC-PR algorithm starts the iteration from the
particle with largest-$p_T$ and, once a stable cone is found, it removes
all particles inside it. Then the procedure starts again with the
hardest remaining particle and go on until no particles are left. This
algorithm is also known as UA1-cone \cite{Arnison:1983gw}, since it was
first introduced and extensively used by the CERN UA1 collaboration.
It is quite easy to see that this algorithm is
IR(collinear)-unsafe. Assume that the hardest particle undergoes
a collinear splitting $p_1 \to p_{1a}, p_{1b}$ with $p_{T,1a}, p_{T,1b}
< p_{T,2} < p_{T,1}$: in this case the IC-PR algorithm would lead to
a different number/configuration of jets, since the $p_T$-ordering has
been modified by the collinear emission.  
The IC-SM algorithm does not rely on any particular ordering
instead. Once all stable cones have been found the prescription is to
{\it merge} a pair of cones if more than a fraction $f$ (typically
$f$=0.5-0.7) of the softer's cone $p_T$ is in common with the harder
one, or to {\it assign} the shared particles to the closer cone. The
{\it split-merge} procedure is repeated until there are no overlapping
cones left.
Unfortunately the IC-SM algorithm is IR(soft)-unsafe as well
\cite{Seymour:1997kj}. Assume that two stable cones are generated
starting from two hard partons whose relative distance is between R and
2R: the addition of an extra soft particle in between would
act as a new seed and the third stable cone would be found, again
leading to a different number/configuration of jets.
A partial solution to this problem was provided by the {\it Midpoint
  Algorithm} \cite{Midpoint}: after all possible jets have been found,
run the algorithm again using additional {\it  midpoint} seeds between
each pair of stable cones. This prescription fixes the IR(soft) issue of
the IC-SM algorithm, since the result is now not dependent on the
presence of an extra soft seed in the overlap region, and has been
adopted as a recommandation for Run II of the Tevatron \cite{Blazey:2000qt}.
Recently \cite{Salam:2007xv} it has been pointed out that, for particular
configurations involving more than two partons, the Midpoint
algorithm is not able to find all stable cones: for exclusive
quantities and/or multi-jet configurations, the midpoint prescription is
still IR(soft)-unsafe.
The IR issue is definitely solved by the introduction of Seedless
algorithms, first proposed in \cite{Blazey:2000qt}. The idea is to
identify all possbile subsets of $N$ particles in an event and, for each
subset $M$, check if the cone defined by the azimuth and rapidity of the
total momentum of $M$ contains other particles outside the subset: if
this is not the case, then $M$ defines a stable cone. With this
prescription the jet-finding algorithm is infrared safe at all
perturbative orders: the main drawback is that the clustering time
(${\cal O} (N \times 2^N$)) leads to extremely slow performances for $N >
$ 4-5. The seedless algorithm has been recently improved by the SIS-Cone
(Seedless Infrared Safe Cone) implementation \cite{Salam:2007xv}, in
which the clustering time is sensibly lowered (${\cal O} (N^2 \log N)$,
comparable to Midpoint). The switching from the midpoint to the seedless
cone is expected to have a significant impact only on exclusive
quantities (i.e., jet mass distribution in multi-jet events), while the
impact for inclusive observables should be modest since Midpoint
IR-unsafety only appears at relatively high orders in perturbation theory.

\section{Sequential Recombination algorithms}

The SR algorithm starts with the introduction of a distance
$d_{ij}=min(k_{ti}^{2p},k_{tj}^{2p})\frac{\Delta_{ij}^2}{R^2}$ between
particles $i,j$ (where $\Delta$ is their distance in the $y,\phi$ plane
and $k_T$ their transverse-momentum), and the distance
$d_{iB}=k_{ti}^{2p}$ between the particle $i$ and the beam. If $d_{ij}
< d_{iB}$ then merge $i$ and $j$, otherwise call $i$ a jet and remove
it from the iteration. There are different types of SR algorithms,
depending on the value of the integer $p$ in the definition of the
distances: $p=1$ identifies the {\it inclusive-$k_T$} algorithm
\cite{Catani:1993hr,Ellis:1993tq}, $p=0$ defines the {\it
  Cambridge-Aachen} algorithm \cite{Dokshitzer:1997in,Wobisch:1998wt},
while for $p=-1$ we have the recently proposed {\it anti-$k_T$}
algorithm \cite{Cacciari:2008gp}. 
All the SR prescriptions are rigorously IR-safe: any soft parton
is first merged with the closest hard parton and only at this point the
decision about the merging of two jets is taken, depending exclusively
on their opening angle. Moreover, there are no more overlap
problems, since any parton is inequivocally assigned to only one jet.
A very fast implementation (clustering time $\sim {\cal O} (N \log N)$) of
all the above SR algorithms is available (FastJet
\cite{Cacciari:2005hq}): it also includes an interface for the
algorithms belonging to the IC class. Another public code providing
access to both SR and IC algorithms is SpartyJet \cite{SpartyJet}.

\section{Summary}

The issue of IR-safety of a jet algorithm should be seriously taken into
account, since multi-jet configurations are sensitive to it and will be
far more widespread at the LHC than at previous colliders. In addition,
without an IR-safe prescription, it would not be possible to fully
exploit the results provided by the theoretical community involved in
NLO multi-leg calculations. Several {\it fast} and {\it safe} algorithms
(SIS-Cone and the SR class) are now available in public packages, but no
definite advantages for a particular algorithm over the others have been
found up to now: the use of different prescriptions for physics analysis
and a continuous cross-checking of results is thus recommended
especially for events with high jet-multiplicity at the LHC.


\begin{thebibliography}{999}

\bibitem{Ellis:1989vm}
  S.~D.~Ellis, Z.~Kunszt and D.~E.~Soper,
  Phys.\ Rev.\  D {\bf 40}, 2188 (1989).

\bibitem{Huth:1990mi}
  J.~E.~Huth {\it et al.},
  Fermilab-Conf-90-249-E (1990).

\bibitem{Ellis:2007ib}
  S.~D.~Ellis, J.~Huston, K.~Hatakeyama, P.~Loch and M.~Tonnesmann,
  Prog.\ Part.\ Nucl.\ Phys.\  {\bf 60}, 484 (2008), 
  and references therein.

\bibitem{Buttar:2008jx}
  C.~Buttar {\it et al.},
  arXiv:0803.0678 [hep-ph], and references therein.

\bibitem{Arnison:1983rn}
  G.~Arnison {\it et al.}  [UA1 Collaboration],
  Phys.\ Lett.\  B {\bf 123}, 115 (1983).

\bibitem{Arnison:1983gw}
  G.~Arnison {\it et al.}  [UA1 Collaboration],
  Phys.\ Lett.\  B {\bf 132}, 214 (1983).

\bibitem{Seymour:1997kj}
  M.~H.~Seymour,
  Nucl.\ Phys.\  B {\bf 513}, 269 (1998).

\bibitem{Midpoint}
  S.~D.~Ellis,
  private communication to the OPAL Collaboration;
  D.~E.~Soper and H.~C.~Yang, 
  private communication to the OPAL Collaboration;
  L.~A.~del~Pozo, 
  University of Cambridge PhD Thesis, RALT-002, (1993);
  R.~Akers {\it et al.},
  OPAL Collaboration, Z. \ Phys. \ C {\bf 63}, 197 (1994).

\bibitem{Blazey:2000qt}
  G.~C.~Blazey {\it et al.},
  arXiv:hep-ex/0005012.

\bibitem{Salam:2007xv}
  G.~P.~Salam and G.~Soyez,
  JHEP {\bf 0705}, 086 (2007).

\bibitem{Catani:1993hr}
  S.~Catani, Y.~L.~Dokshitzer, M.~H.~Seymour and B.~R.~Webber,
  Nucl.\ Phys.\  B {\bf 406}, 187 (1993).

\bibitem{Ellis:1993tq}
  S.~D.~Ellis and D.~E.~Soper,
  Phys.\ Rev.\  D {\bf 48}, 3160 (1993).

\bibitem{Dokshitzer:1997in}
  Y.~L.~Dokshitzer, G.~D.~Leder, S.~Moretti and B.~R.~Webber,
  JHEP {\bf 9708}, 001 (1997).

\bibitem{Wobisch:1998wt}
  M.~Wobisch and T.~Wengler,
  arXiv:hep-ph/9907280.

\bibitem{Cacciari:2008gp}
  M.~Cacciari, G.~P.~Salam and G.~Soyez,
  JHEP {\bf 0804}, 063 (2008).

\bibitem{Cacciari:2005hq}
  M.~Cacciari and G.~P.~Salam,
  Phys.\ Lett.\  B {\bf 641}, 57 (2006);
  http://www.lpthe.jussieu.fr/$\sim$salam/fastjet/

\bibitem{SpartyJet}
  http://www.pa.msu.edu/$\sim$huston/SpartyJet/SpartyJet.html.

\end{thebibliography}
\end{document}